# Observation of bands with $d_{xy}$ orbital character near the Fermi level in NdFeAs$_{1-x}$P$_x$O$_{0.9}$F$_{0.1}$ using angle-resolved photoemission spectroscopy

Z. H. Tin,[1] T. Adachi,[1] A. Takemori,[1] K. Yoshino,[1] N. Katayama,[1] S. Miyasaka,[1] S. Ideta,[2,3] K. Tanaka,[3,4] and S. Tajima[1]

[1]*Department of Physics, Osaka University, Osaka 560-0043, Japan*
[2]*Hiroshima Synchrotron Radiation Center, Hiroshima University, Higashi-Hiroshima, 739-0046, Japan*
[3]*UVSOR Synchrotron Facility, Institute for Molecular Science, Okazaki 444-8585, Japan*
[4]*School of Physical Sciences, The Graduate University for Advanced Studies (SOKENDAI), Okazaki 444-8585, Japan*

We studied the band structure of NdFeAs$_{1-x}$P$_x$O$_{0.9}$F$_{0.1}$ ($x$ = 0, 0.2, 0.4 and 0.6) using angle-resolved photoemission spectroscopy (ARPES) measurements. Two of the hole bands, α$_1$ ($d_{xz}$) and α$_3$ ($d_{z^2}$), were observed at the Brillouin zone center in the *P*-polarized light configuration, while the other two hole bands, α$_2$ ($d_{yz}$) and γ ($d_{xy}$), were observed in the *S*-polarized alternative. The observed γ band shifts downwards as $x$ increases, which is consistent with the theoretical prediction for the change in bond angle of As/P-Fe-As/P. Furthermore, a small amount of the $d_{xy}$ orbital component was observed at the same binding energy as that of the top of the α$_1$ band, thus indicating the band reconstruction of the originally degenerate α$_1$ and α$_2$ ($d_{xz}/d_{yz}$) bands by the unoccupied $d_{xy}$ band. The change in the energy level of the α$_1$ band top with $d_{xy}$ orbital character is accompanied by a $T_c$ upturn at $0.2 < x < 0.4$. The $T_c$ continues to increase as the α$_1$ band shifts downward, crossing the Fermi level. The incipient band with the $d_{xy}$ orbital character on its top could be an important ingredient for high $T_c$ 1111-type iron-based superconductors.

## I. INTRODUCTION

The discovery of iron-based superconductors (IBSs) [1] has garnered significant attention and provided a new platform to study novel mechanisms of superconductivity, in addition to unconventional copper-oxide superconductors. Although various types of IBSs with different crystal structures have been discovered, they all share the common structure of the Fe*Pn* (*Pn* = pnictogen) layer, which functions as the conducting layer. In the case of 1111-type IBSs, such as *R*FeAsO (*R* = rare-earth elements), the structure comprises an alternating stack of insulating [*R*O]$^+$ and conducting [FeAs]$^-$ layers. The parent compound, i.e., *R*FeAsO, exhibits a structural phase transition accompanied by a magnetic transition into the antiferromagnetic order (AFM1) with decreasing temperature [2,3]. The substitution of oxygen with fluorine (aliovalent doping), which introduces electrons into the FeAs conducting layer, suppresses the AFM1 and induces superconductivity (SC1).

In addition to electron doping, superconductivity can be achieved by substituting arsenic with phosphorus, e.g., via isovalent doping. The results of the crystal structure analyses indicate that the pnictogen height from Fe plane $h_{Pn}$ and the As/P–Fe–As/P bond angle *θ* change monotonically with phosphorus doping [4–8]. However, the electronic properties, including the superconducting transition temperature ($T_c$), do not change linearly with phosphorus-doping. For example, in a system with 5% electron doping, LaFeAs$_{1-x}$P$_x$O$_{0.95}$F$_{0.05}$, the $T_c$ exhibits two-dome structures called SC1 and SC2 with increasing *x*. The non-monotonic change in $T_c$ indicates the possibility of two different electronic states in the SC1 and SC2 regions [4,6–9]. The presence of two superconducting phases is further supported by the two different types of antiferromagnetic orders, namely, AFM1 and AFM2, in the zero-electron doping system [10]. In other words, AFM1(AFM2) is transitioned to SC1(SC2) upon electron doping. Our previous transport measurement and angle-resolved photoemission spectroscopy (ARPES) measurements of NdFeAs$_{1-x}$P$_x$O$_{0.9}$F$_{0.1}$ revealed that the switching point of these two electronic states is approximately $x$ = 0.2 [4,8,11].

Band calculation predicts the evolution of a two-dimensional (2D) $d_{xy}$ into a three-dimensional $d_{z^2}$ orbital hole Fermi surface (FS) with phosphorus doping [12]. Therefore, it is natural to attribute the non-monotonic electronic change with *x* to this FS change [13]. In the spin fluctuation model, the 2D $d_{xy}$ hole pocket is crucial in the high $T_c$ superconductivity with the full-gapped s± wave symmetry [14–16]. The importance of the $d_{xy}$ orbital in enhancing $T_c$ is also predicted in the strong electron correlation approach [17–23]. However, the $d_{xy}$ hole pocket has never been experimentally observed to date in 1111-type IBSs.

In this study, by focusing on the low-binding-energy region, we performed an ARPES study on NdFeAs$_{1-x}$P$_x$O$_{0.9}$F$_{0.1}$ ($x$ = 0, 0.2, 0.4 and 0.6), and then successfully observed the two states with the $d_{xy}$ orbital characteristic. One state is the γ hole band at 52–80 meV below the Fermi level ($E_F$) at the center of the folded Brillouin zone $(k_x, k_y) = (0,0)$. The other band with the $d_{xy}$ orbital character is located just below $E_F$ and exhibits similar trends with the α$_1$ band, thus suggesting the change in orbital character of the α$_1$ band by the band reconstruction at the zone center.

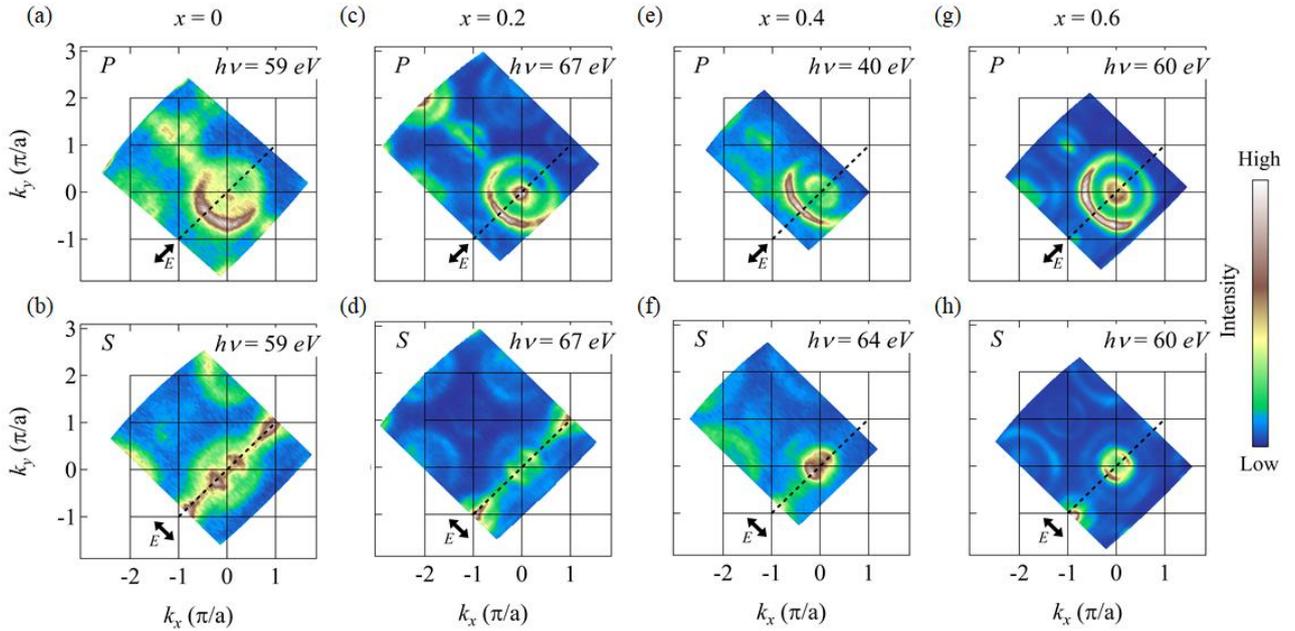

FIG. 1. In-plane FS intensity mapping of NdFeAs$_{1-x}$P$_x$O$_{0.9}$F$_{0.1}$. The dotted black line marks the direction of the mirror plane of the system. The double-sided arrow denotes the direction of the polarization vector of incident light. (a,b) FS mapping of $x = 0$ taken by incident light with $h\nu = 59$ eV, (c,d) $x = 0.2$ taken by incident light with $h\nu = 67$ eV, (e,f) $x = 0.4$ taken by incident light with $h\nu = 40$ and 64 eV, (g,h) $x = 0.6$ taken by incident light with $h\nu = 60$ eV. Upper ((a), (c), (e), and (g)) and lower ((b), (d), (f), and (h)) panels present the results in $P$- and $S$-polarization configurations, respectively.

## II. EXPERIMENTALS

Single crystals of NdFeAs$_{1-x}$P$_x$O$_{0.9}$F$_{0.1}$ were grown under high-pressure, as reported in a previous study [11]. The actual P content $x$ was verified to be the same as the nominal within the error of $\pm\,0.03$ via energy-dispersive X-ray spectroscopy, while the F content $y$ estimated using an electron probe microanalyzer was approximately 0.05. In this study, we will adopt the nominal concentrations of P and F as $x$ and $y$, respectively. The $T_c$ values of NdFeAs$_{1-x}$P$_x$O$_{0.9}$F$_{0.1}$ ($x = 0, 0.2, 0.4$ and $0.6$) determined by magnetic susceptibility measurements in a magnetic field of 10 Oe were 43 K, 24 K, 16 K and 11 K, respectively. The ARPES measurements were performed at the BL 5U and 7U of the UVSOR-III Synchrotron at the Institute for Molecular Science [24]. The energy and angular resolution were approximately ~12 meV and ~0.2°, respectively. Linear polarized incident photons and the MBS A1 analyzer were used for all the measurements. All samples were cleaved in-situ just above $T_c$ in an ultrahigh vacuum of approximately $5 \times 10^{-9}$ Pa, while the ARPES spectra were measured at 10 K. The calibration of the Fermi level ($E_F$) was achieved by referring to the spectra of gold, which is electrically contacted with a sample.

The top-most layer of the cleaved surface is either the NdO or FeAs layer owing to the alternating layer nature of 1111 system. All the ARPES results on the 1111 system exhibited a large hole FS, $\alpha_3$. C. Liu et al. claimed that the surface layer will create a large hole FS from the charge transfer scenario, regardless of the type of surface layer [25]. From the surface cleavage study, H. Eschrig et al. asserted that only the FeAs surface layer produces a large hole FS [26]. Hence, the top plane exposed to the vacuum is likely to be the FeAs terminated layer, which is consistent with the results obtained by Yang et al. [27].

The photoemission intensity is approximately proportional to the combination of the matrix element, Fermi–Dirac distribution function, and single-particle spectral function. The matrix element effect allows us to determine the orbital character of the particular band in the specific experimental configuration. In all the samples, $x$ and $y$ axes are parallel to the Fe-Fe bond direction within the plane. In the $P$ ($S$) polarization geometry, the vector of incidents photons' electric field is parallel (perpendicular) to the mirror plane defined by the analyzer slit parallel to the $x$ axis and normal vector ($z$ axis) of the cleaved sample surface. The propagator vectors of the incident photons are located in the $xz$ mirror plane [24]. By manipulating the polarization of the incident photons relative to the mirror plane, we can then resolve odd and even symmetry of the orbital characters [28–31]. In summary, horizontally polarized light ($P$) can detect $d_{xz}$, $d_{x^2-y^2}$, and $d_{z^2}$ orbital characters, while vertically polarized light ($S$) can detect $d_{yz}$ and $d_{xy}$ orbital characters, as summarized in Table 1.

Because the analyzer slit is parallel to $x$ axis, the FSs and bands with the $d_{xz}$ and $d_{yz}$ orbital characters can be distinguished from the perspective of an experimental setting by adopting $P$ and $S$-polarized light, respectively. However, in the entire P-doping and temperature regions, the present system, NdFeAs$_{1-x}$P$_x$O$_{0.9}$F$_{0.1}$ has a tetragonal crystal structure, where the $d_{xz}$ and $d_{yz}$ orbital characters of FSs and electronic bands are essentially equivalent. In latter sections, these orbital characters are distinguished, when we present and discuss the ARPES results. Nevertheless, in the discussion part for the general electronic properties, the $d_{xz}$ and $d_{yz}$ orbital characters are considered as equivalent.

Table 1 Allowed symmetries with respect to the $xz$ mirror plane.

| Polarization configuration | $d_{xz}$ | $d_{yz}$ | $d_{xy}$ | $d_{x^2-y^2}$ | $d_{z^2}$ |
|---|---|---|---|---|---|
| S | | ✓ | ✓ | | |
| P | ✓ | | | ✓ | ✓ |

## III. RESULTS AND DISCUSSION

The in-plane FS mapping of NdFeAs$_{1-x}$P$_x$O$_{0.9}$F$_{0.1}$ ($x = 0$, 0.2, 0.4 and 0.6) with two different polarizations is illustrated in Fig. 1, where the intensities of the energy distribution curves (EDCs) were integrated within the energy region of ± 10 meV with respect to $E_F$. Note that the folded Brillouin zone was adopted in our data, instead of the unfolded Brillouin zone used in the theoretical approach [14–16]. First, in the *P*-polarization, we observed one large hole FS ($\alpha_3$) centered at (0,0) for all $x$ values. Extra inner hole FS can be observed for $x \geq 0.4$. A propeller-like electron FS was observed at the zone corner ($\pi$, $\pi$), especially for $x = 0$ and 0.2, but absent in $x = 0.4$ and 0.6. In the *S*-polarization, an inner hole FS ($\alpha_2$) was observed at (0,0) for all $x$ values [32,33].

To determine the high-symmetry point, we measured the $k_z$ dependence of the FS topology by varying the photon energy of incident photons in both the *P*- and *S*-configurations at the zone center. The obtaind results are consistent with the previous finding that the FS topology remains unchanged along the $k_z$ direction in the 1111 system [25]. However, we identified that two bands below the $E_F$ exhibit periodic appearance and disappearance in vertically polarized light configuration within a wide photon energy region in the beamlines, BL 7U and 5U covering two Brillouin zones. This behavior is most likely intrinsic in NdFeAs$_{1-x}$P$_x$O$_{0.9}$F$_{0.1}$. These bands emerge when the photon energies are approximately 18 eV, 34 eV, and 56 eV, which correspond to the same $k_z$ position in the periodic Brillouin zone. We deduced that these energies indicate the high symmetry points, similar to the report presented by Yang et al. [27] Here, we fixed the photon energies of our measurement at approximately 58 eV in the BL5U beamlines, while the ~18 eV and ~36 eV energies were fixed in the BL 7U beamlines.

As mentioned above, in the in-plane FS mapping, only two hole FSs ($\alpha_2$ and $\alpha_3$) were observed in NdFeAsO$_{0.9}$F$_{0.1}$ ($x = 0$), although the band calculation predicted three hole FSs [34]. To clarify this contradiction, we observed an energy-momentum (*E-k*) cut along the high-symmetry direction (dotted black line in Fig. 1(a)). The results obtained from the *P*-polarization for $x = 0$ is presented in Fig. 2(a), while that of the *S*-polarization data is shown in Fig. 2(d). The second derivatives of the intensity with respect to momentum and energy are illustrated in Figs. 2(b, e) and 2(c,f). To trace the band dispersion, we applied four Lorentzian fittings on the momentum distribution curves (MDCs). The obtained energy dependence of the MDC peak position can then be fitted with a parabola curve. The obtained parabola curves were plotted on the original and second derivative intensity map, as illustrated in Figs. 2(a–i). In the *P*-polarization (Figs. 2(a–c)), two hole bands can be observed. The large hole band $\alpha_3$ in the *P*-polarization (denoted by the black dotted line) is assigned to the $d_{z^2}$ band, while the smaller hole band $\alpha_1$, whose band top is approximately 18 meV below $E_F$, should exhibit an orbital character of $d_{xz}$ because it is only visible in the *P*-polarization. In contrast, in the *S*-polarization (Figs. 2(d–f)), we observe the $d_{yz}$ ($\alpha_2$) band, which is commonly found in the ARPES of IBSs. In addition, another hole band ($\gamma$) is observed with its band top at 52 meV below the $E_F$.

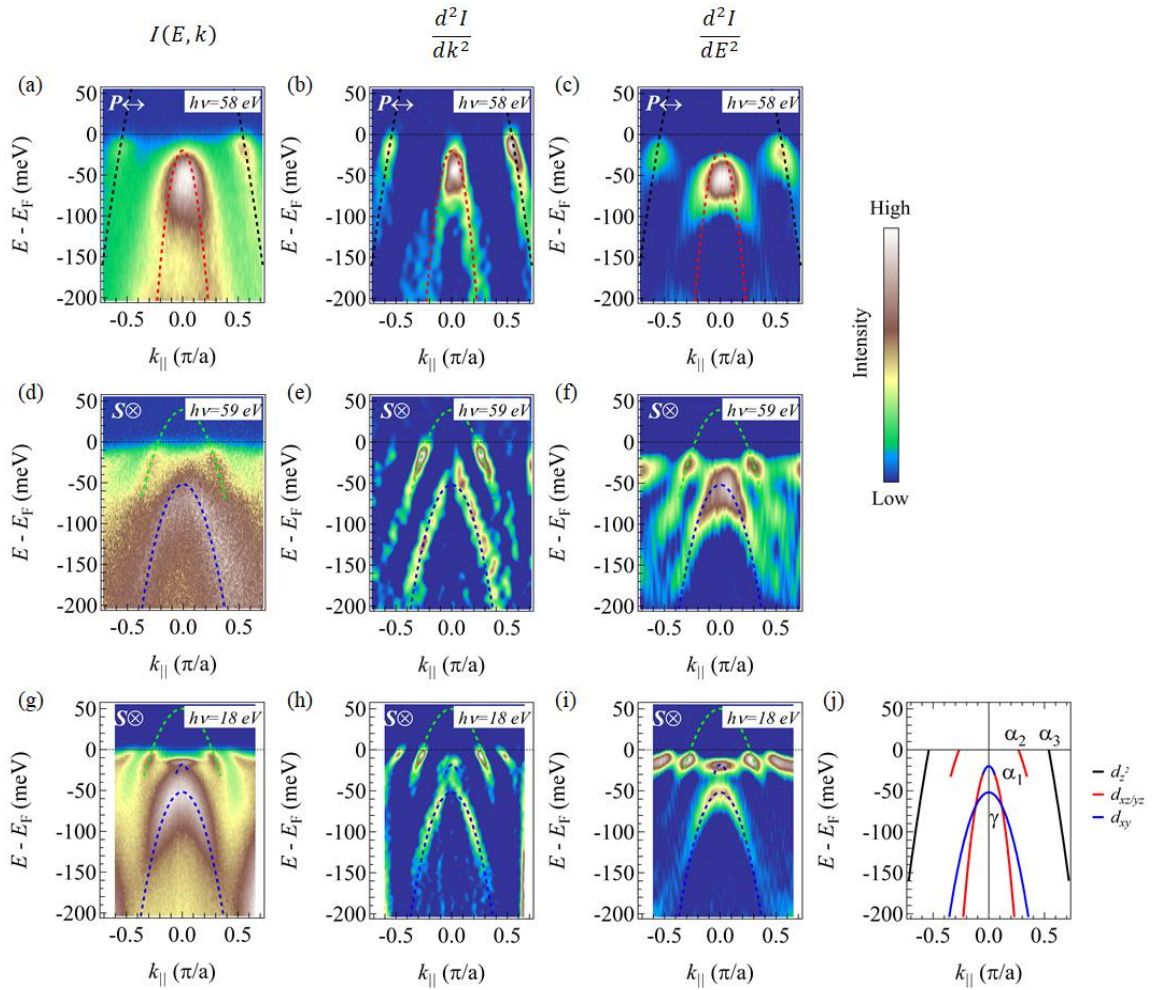

FIG. 2. ARPES intensity plots of NdFeAsO$_{0.9}$F$_{0.1}$ taken around the Brillouin zone center. a) The intensity plot and (b,c) their second derivative plots with respect to *k* and *E*, respectively, taken by incident light with *hν* = 58 eV in the *P*-polarization configuration. The black and red dotted lines represent the $d_{z^2}$ and $d_{xz}$ bands ($\alpha_3$ and $\alpha_1$). (d) The intensity plot and (e,f) their second derivative plots with respect to *k* and *E*, respectively,, taken by incident light with *hν* = 58 eV in the *S*-polarization configuration. The green and blue dotted lines depict the $d_{yz}$ and $d_{xy}$ bands ($\alpha_2$ and $\gamma$). (g) The intensity plot and (h,i) their second derivative plots with respect to *k* and *E*, respectively, taken by incident light with *hν* = 18 eV in the *S*-polarization configuration. The green and blue dotted lines represent the $d_{yz}$($\alpha_2$) and $d_{xy}$ bands (top of $\alpha_1$ and $\gamma$). (j) Summarized band dispersion and the orbital character of NdFeAsO$_{0.9}$F$_{0.1}$. The general discussion of electronic bands, the same color (red) line depicts the $d_{xz}$ and $d_{yz}$ bands, while the black and blue ones depicts the $d_{z^2}$ and $d_{xy}$ bands in this panel (j), respectively.

To validate the intrinsicality of these bands, we repeated the same measurement with lower photon energy (18 eV) at the zone center, and at another beamline (BL7U). Figs. 2(g–i) shows the original intensity $I(E,k)$, the second derivatives $\frac{d^2I}{dk^2}$, and $\frac{d^2I}{dE^2}$ in the *S*-polarization, respectively. The high-resolution measurement at $h\nu = 18$ eV reproduced the band dispersion of Figs. 2(c–f). The band depicted by the blue dotted line is only visible in the *S*-polarization, and thus primarily originates from the $d_{xy}$ orbital. We assigned this band as the previously missing $\gamma$ band, which should cross $E_F$ according to the theoretical calculation [34]. Surprisingly, in the present ARPES results, we observed a small flat band, which is located at around $k_\parallel = 0$, at $E - E_F = -18$ meV. This flat band remains almost unchanged, after dividing the ARPES data by the Fermi-Dirac function, indicating that it is a real feature of the electronic band structure [35]. (In addition to $x = 0$, we have confirmed that the flat band is clearly observed in the ARPES data divided by the Fermi-Dirac function in $x = 0.2$ and 0.4. The results of ARPES for $x = 0.2$ and 0.4 are shown later.) We speculate that this flat band originates from the $\alpha_1$ band, as they are in the same energy region. Here, we want to stress that this flat band does not emerge from the $d_{yz}$ orbital part of $\alpha_1$. (Remember that both the orbitals of $\alpha_1$ and $\alpha_2$ are changing between $d_{xz}$ and $d_{yz}$ with every 90°.) If this flat dispersion spanning from $-0.1 \leq k_\parallel \leq 0.1$ originated from the $d_{yz}$ orbital part of $\alpha_1$, the ARPES measurement at 18 eV (40 eV) must be misaligned from the center (0,0) for approximately 2° (1.5°). Such misalignment will lead to the 40 meV downshift of a local maximum of the hole band. However, such a huge downshift is not observed. Furthermore, we have confirmed that our data cut through the center of this flat dispersion, based on the FS mapping intensity analysis at $E - E_F = -15$ meV. We argue that this flat dispersion has an orbital character of $d_{xy}$ and suggests a switch of orbital character from $d_{xy}$ to $d_{xz}$, as the momentum moves away from $k_\parallel = 0$ of the $\alpha_1$ band. The band structure and orbital character are summarized in Fig. 2(j).

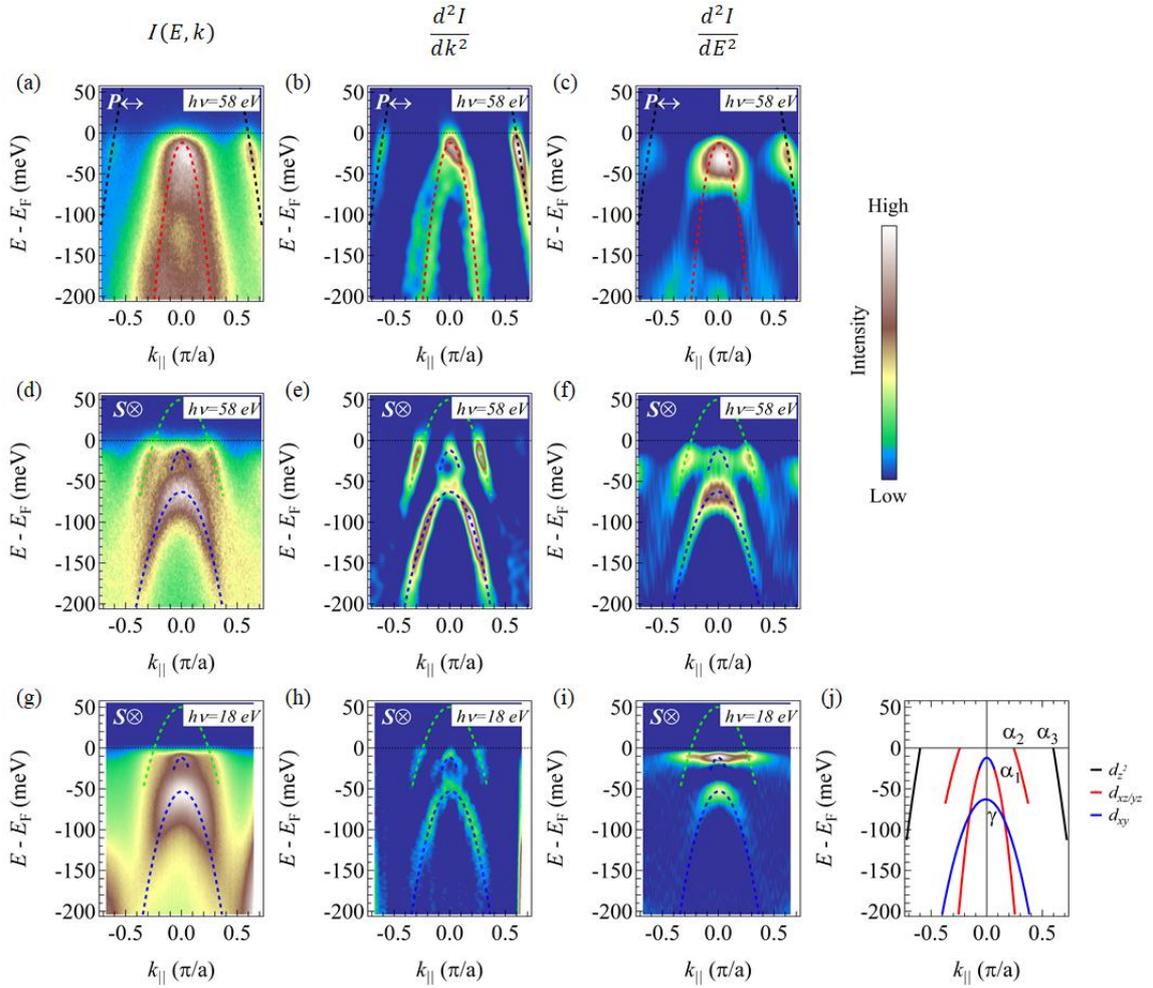

FIG. 3. ARPES data of NdFeAs$_{0.8}$P$_{0.2}$O$_{0.9}$F$_{0.1}$ around the Brillouin zone center. (a) The intensity plot and (b,c) their second derivative plots with respect to $k$ and $E$, respectively, taken by incident light with $h\nu = 58$ eV in the *P*-polarization configuration. The black and red dotted lines indicate the $d_{z^2}$ and $d_{xz}$ bands ($\alpha_3$ and $\alpha_1$). (d) The intensity plot and (e,f) their second derivative plots with respect to $k$ and $E$, respectively, taken by incident light with $h\nu = 58$ eV in the S-polarization configuration. The green and blue dotted lines indicate the $d_{yz}$ and $d_{xy}$ bands ($\alpha_2$, top of $\alpha_1$ and $\gamma$. (g) The intensity plot and (h,i) their second derivative plots with respect to $k$ and $E$, respectively, taken by incident light with $h\nu = 18$ eV in the *S*-polarization configuration. (j) Summarized band dispersion and the orbital character of NdFeAs$_{0.8}$P$_{0.2}$O$_{0.9}$F$_{0.1}$. In this panel (j) for the general discussion of electronic bands, the same color (red) line indicates the $d_{xz}$ and $d_{yz}$ bands, while the black and blue ones indicate the $d_{z^2}$ and $d_{xy}$ bands.

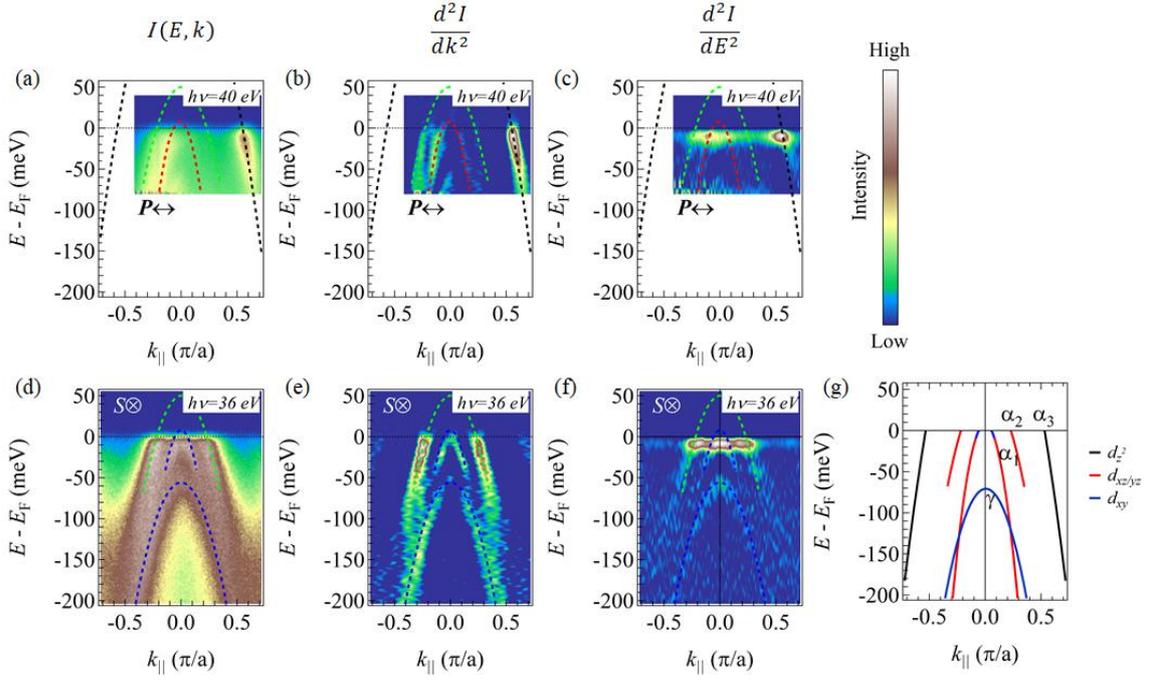

FIG. 4. ARPES intensity plots of NdFeAs$_{0.6}$P$_{0.4}$O$_{0.9}$F$_{0.1}$ around the Brillouin zone center. (a) The intensity plot and (b,c) their second derivative plots with respect to $k$ and $E$, respectively, taken by incident light with $h\nu = 40$ eV in the $P$-polarization configuration. (d) The intensity data and (e,f) their second derivative plots with respect to $k$ and $E$, respectively, taken by incident light with $h\nu = 36$ eV in the $S$-polarization configuration. (g) Summarized band dispersion and the orbital character of NdFeAs$_{0.6}$P$_{0.4}$O$_{0.9}$F$_{0.1}$.

Fig. 3 presents the ARPES results for $x = 0.2$. In the $P$-polarization (Figs. 3(a–c)), we can clearly observe the $\alpha_1$, and $\alpha_3$ bands, similar to the $x = 0$ case in Fig. 2. Compared with the results of $x = 0$, the $\alpha_1$ band is closer to the $E_F$. The top of the $\alpha_1$ band is approximately 12 meV below $E_F$. In the $S$-polarization (Figs. 3(d–f)), the typical $d_{yz}$ $\alpha_2$ band is observed together with the $\gamma$ band with the $d_{xy}$ orbital character. The band top position of the $\gamma$ band is approximately 63 meV below $E_F$. Moreover, a flat band can be observed near $E_F$. A similar band structure was obtained with the lower photon energy (18eV) (Figs. 3(g–i)). Again, the flat band visible in the $S$-polarization exhibits the same energy (-12 meV) as that of the top of the $\alpha_1$ band observed in the $P$-polarization in Figs. 3(a, b). This result further convinces us that the $d_{xy}$ orbital character is on the band top of the original $d_{xz}$ $\alpha_1$ band. We summarize the band structure of NdFeAs$_{0.8}$P$_{0.2}$O$_{0.9}$F$_{0.1}$ with its corresponding orbital character in Fig. 3(j).

Figs. 4 (a–c) present the results of ARPES measurements in the $P$-polarization configuration for $x = 0.4$. As $x$ value increases up to $x = 0.4$, the $\alpha_1$ band shifts up, crosses the $E_F$, and then forms a very small FS, which is already presented in Fig. 1. Surprisingly, the $\alpha_2$ band is observed in $P$-polarization. Recall that $\alpha_1$ ($\alpha_2$) bands can be observed only in $P$- ($S$-) polarization owing to the matrix element effect of $d_{xz}(d_{yz})$ orbital character. This may raise doubts in that the observation of both the $\alpha_1$ and $\alpha_2$ bands in $P$-polarization is due to off center measurement. However, the center cut was determined from the 2D FS plot, which is a series collection of energy-momentum cut with 1° intervals by a tilt goniometer within a 0.5° accuracy. The results of the FS mapping in Fig. 1 have clearly presented almost no misalignment in the present experiments. Moreover, the in-plane FS mappings in the bottom panels of Fig.1 distinctly demonstrate the evolution of the inner hole FS from the round bracket to the circular shape at the zone center. Therefore, we argued that the observations of both the $\alpha_1$ and $\alpha_2$ bands in $P$-polarization in $x = 0.4$ is intrinsic and can be explained by the gradual change of $d_{xz}/d_{yz}$ to $d_{XZ}/d_{YZ}$ orbital character in this band, which has been theoretically predicted [13]. Here, $X$ and $Y$ axes are parallel to $x$-$y$ and $x+y$ directions, respectively. In addition, the $\alpha_2$, $\gamma$ bands and the additional band near $E_F$ (possibly related to the $\alpha_1$ band) can be observed in $S$-polarization, as illustrated in Figs. 4(d–f). The $\gamma$ band has a binding energy similar to that of $x = 0.2$. The flat $d_{xy}$ band observed in $x = 0$ and 0.2 is more dispersive in $x = 0.4$ and follows the trend of $\alpha_1$ band, crossing the $E_F$.

In the $P$-polarization of $x = 0.6$, the $\alpha_1$ band shifts up further and forms a larger FS, as illustrated in Figs. 5(a–c). Similar to $x = 0.4$, the $\alpha_2$ band can be observed in the $P$-polarization with a photon energy of 60 eV in $x = 0.6$, thus suggesting the change in orbital character from $d_{xz}/d_{yz}$ to $d_{XZ}/d_{YZ}$. This switching from $d_{xz}/d_{yz}$ to $d_{XZ}/d_{YZ}$ orbital character is replicated using different pieces of single crystals with a photon energy of 40 eV, as illustrated in Figs. 5(g–i). In fact, the thick inner hole FS observed in Fig. 1(g) originates from the contributions of both $\alpha_1$ and $\alpha_2$ FSs. In addition to the common $\alpha_1$, $\alpha_2$, and $\alpha_3$ bands, an additional band (marked as a white dotted line in Figs. 5(a–c)) is observed. Although this additional band has a similar band top energy with the $\gamma$ band, the band dispersion completely differs from the $\gamma$ band. This additional band may be assigned to be the band with the $d_{z^2}$ orbital, as theoretical studies have demonstrated that the $d_{z^2}$ band in $x = 1.0$ exhibits a higher energy level than that in $x = 0$, i.e., $d_{z^2}$ band is closer to $E_F$ for the P end member [15,16]. In the $S$-polarization (Figs. 5(d–f)), only the $\alpha_2$ and $\gamma$ bands can be observed clearly. The $\gamma$ band for $x = 0.6$ is located in the lower energy region than that for $x \leq 0.4$, and the energy level of the band top reaches $E - E_F = -80$ meV. The second derivative plots reveal a hidden dispersion, just above the $\gamma$ bands. This weak dispersion is a hyperbola (nearly linear dispersion as observed in the Dirac-cone type), which differs from the bulk band discussed so far. It is speculated to be a surface state of $\gamma$ bands ($d_{xy}$) as it appears to be connected to the $\gamma$ bands in $S$-polarization configuration. In contrast, the $\alpha_1$ band shifts up with increasing $x$, and the band top is far from $E_F$ in $x = 0.6$, as illustrated in Fig. 5. This energy level increment of $\alpha_1$ band may hide its $d_{xy}$ orbital character, which is observed only around the $\alpha_1$ band top in the samples with $x \leq 0.4$.

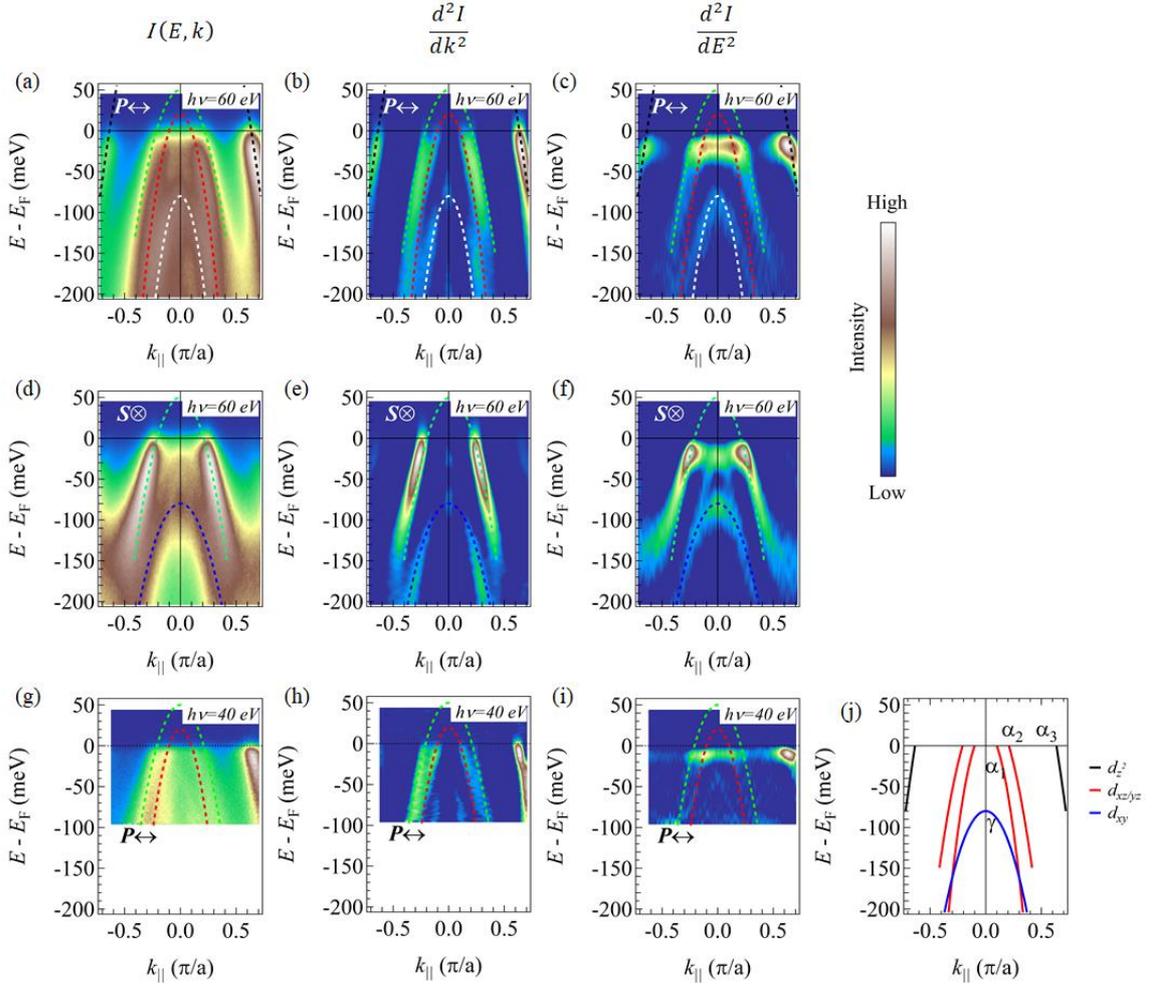

FIG. 5. ARPES intensity plots of NdFeAs$_{0.4}$P$_{0.6}$O$_{0.9}$F$_{0.1}$ around the Brillouin zone center. (a) The intensity plot and (b,c) their second derivative plots with respect to $k$ and $E$, respectively, taken by incident light with $h\nu = 60$ eV in the $P$-polarization configuration. (d) The intensity data and (e,f) their second derivative plots with respect to $k$ and $E$, respectively, taken by incident light with $h\nu = 60$ eV in the $S$-polarization configuration. (g) The intensity data and (h,i) their second derivative plots with respect to $k$ and $E$, respectively, taken by incident light with $h\nu = 40$ eV in the $P$-polarization configuration. (j) Summarized band dispersions and the orbital characters of NdFeAs$_{0.4}$P$_{0.6}$O$_{0.9}$F$_{0.1}$.

Based on the present ARPES results, we summarize and present the schematic plot of the $x$ dependence of the NdFeAs$_{1-x}$P$_x$O$_{0.9}$F$_{0.1}$ band structure in Fig 6. Upon P doping, the $\alpha_1$ band continuously shifts up and eventually cross $E_F$ between $x = 0.2$ and 0.4. The appearance of the additional FS, which was induced by the shift of the $\alpha_1$ band top up to $E_F$, was already verified in this $x$ region in our previous study [11]. It is speculated that, further increasing the P content will shift the $\alpha_1$ band upwards continuously, and eventually becomes degenerate with the $\alpha_2$ band. On the other hand, the $\gamma$ band shifts downwards to lower energy regions with increasing P-doping level.

It is well known that the surface reconstruction of the 1111 system after cleaving (owing to its polar charge surface properties), is the primary reason for the inconsistent results between the surface-sensitive ARPES and band calculation. Several studies have been conducted to disentangle the surface and bulk band structures [25,27,36,37]. The large $d_{z^2}$ $\alpha_3$ hole FS was reported to be extrinsic owing to intensity suppression upon thermal cycle and Na/K dosing. However, the inner hole FS ($\alpha_2$) and the propeller-like electron FS were quite controversial. Yang et al. claimed that the $\alpha_2$ FS is bulk-related while the propeller-like electron FS is surface-related [27]. Conversely, Charnukha et al. argued the opposite [36]. Here, we believe that the $\alpha_2$ band is a bulk band because it is only visible in the S-polarizations ($d_{yz}$ part of $d_{xz}/d_{yz}$) in $x = 0$ and 0.2, but appears in the P-polarization when $x \geq 0.4$ ($d_{yz} \rightarrow d_{YZ}$, where $d_{YZ}$ is visible in both P- and S-polarizations). This switching of orbital character from $d_{xz/yz}$ to $d_{XZ/YZ}$ is due to the de-hybridization, which is predicted in the calculation [13]. A clear polarization dependence and a systematic $x$-dependence in our data indicate that the present observations are bulk properties, although there may be some surface effects in the ARPES experiment such as the energy shift due to the surface compression [26], which explained the large $d_{z^2}$ $\alpha_3$ band.

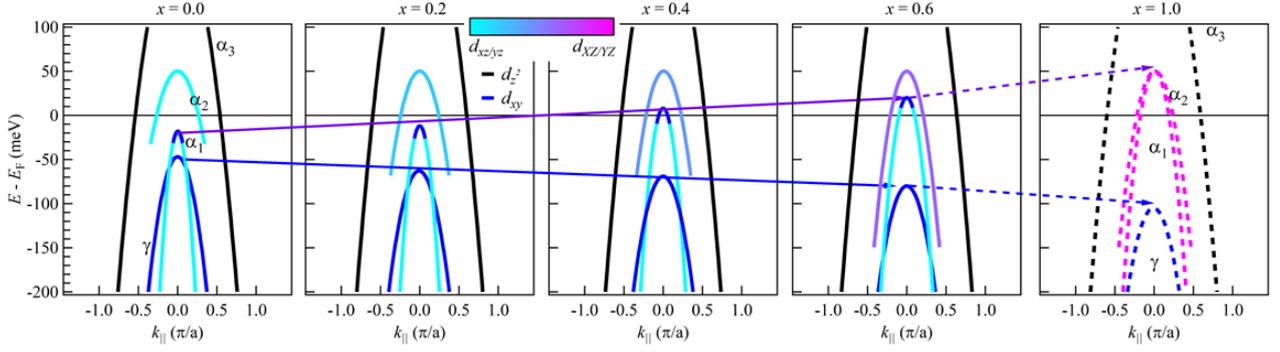

FIG. 6. Schematic illustration of the band structure evolution at the Brillouin zone center with P-doping in NdFeAs$_{1-x}$P$_x$O$_{0.9}$F$_{0.1}$. The bands with $d_{xy}$ orbital character ($\gamma$ band and the top of $\alpha_1$ band) are represented by the solid blue lines. The $\alpha_3$ bands with $d_{z^2}$ orbital character is depicted by the solid black lines. With increasing $x$, the orbital characters of $\alpha_1$ and $\alpha_2$ bands are gradually changed from $d_{xz}/d_{yz}$ to $d_{XZ}/d_{YZ}$. The continuous change of orbital character is represented by the color gradation of the $\alpha_1$ and $\alpha_2$ band lines. The speculated band structure for $x = 1.0$ is also shown.

The $\alpha_1$ band below $E_F$ was also observed in other high-$T_c$ 1111 systems, i.e., NdFeAsO$_{0.4}$F$_{0.6}$, SmFe$_{0.92}$Co$_{0.08}$AsO, and PrFeAsO$_{0.7}$ [32,33,36]. The low-lying $\alpha_1$ band, namely, incipient band, is possibly a universal feature in the 1111-type IBSs with a small bond angle. The same $\alpha_1$ band was observed to form FS in the normal state of LaFeAsO [27,37]. It is expected that this $\alpha_1$ FS moves from $E_F$ and forms an incipient band upon electron-doping. To date, there is no report claiming this incipient band is surface-related.

However, regarding the small flat band observed in the S-polarization at -18 meV (-12 meV, around $E_F$) for $x = 0$ (0.2, 0.4), there are three reasons to believe that this band originates from the $\alpha_1$ band. First, the flat band is observed in the S-polarization within a confined region of energy and momentum. Second, the $\alpha_1$ ($d_{xz}$) band top is observed in the P-polarization at the same energy as the flat band with the $d_{xy}$ orbital character in the S-polarization. Both the band top of $\alpha_1$ and the flat $d_{xy}$ band shift upwards by the same magnitude (6 meV) from $x = 0$ to 0.2, and then cross the $E_F$ for $x = 0.4$. Third, there is a theoretical prediction of orbital switching at the top of the $\alpha_1$ band if a certain condition suffices.

According to the band calculation, the $\alpha_1$ band is degenerate with $\alpha_2$ at $x = 1.0$, where the bond angle $\theta$ is large. Usui et al. argued that the reduction in the bond angle $\theta$ results in the decrease in nearest-neighbor hopping $t_1$ in the five-band model [34]. Consequently, the upper $d_{xy}$ band above $E_F$ shifts down and touches the degenerate $d_{xz}/d_{yz}$ hole bands (degenerate $\alpha_1$ and $\alpha_2$), which induces the band reconstruction and splitting of these bands. A small amount of $d_{xy}$ orbital character at the band top of the $d_{xz}$ $\alpha_1$ band is the outcome of the band splitting process of the degenerate $d_{xz}/d_{yz}$ hole bands. The present data are in line with this band-splitting scenario. However, there is a slight difference between the calculation and our data, which is the critical bond angle for the occurrence of band splitting. Theoretically, the band splitting occurs when the bond angle $\theta \leq 109°$ [34], while the smallest angle in our Nd-1111 compound is $\theta \sim 111°$ at $x = 0$, based on our previous structural analysis of polycrystalline NdFeAs$_{1-x}$P$_x$O$_{0.9}$F$_{0.1}$ via synchrotron X-ray diffraction [8,11].

In addition to the band splitting of the degenerate $d_{xz}/d_{yz}$ hole bands, the $d_{xy}$ band is theoretically predicted to shift upward with decreasing $\theta$ [34,38,39]. The appearance of the extra $d_{xy}$ hole FS plays is crucial in the superconductivity of the full gap s$\pm$ pairing mechanism in the As-end compound. However, this $d_{xy}$ hole FS has never been reported in the 1111 system. The result obtained from this study capture the possible $d_{xy}$ orbital character band ($\gamma$). However, this $\gamma$ band does not cross the $E_F$ for all $x$ samples in this study, which contradicts band calculation [13,34]. Nevertheless, the band top of $\gamma$ shifts upward with decreasing $x$, namely, decreasing the bond angle $\theta$, which qualitatively agrees well with the theoretical prediction [12,34]. The systematic energy shift of the $\alpha_1$ and $\gamma$ bands tops with P-doping is a clear indicator of bulk electronic structure properties.

Our results capture the systematic evolution of orbital character and the band structure in NdFeAs$_{1-x}$P$_x$O$_{0.9}$F$_{0.1}$. As summarized in Figs. 6 and 7, the $\gamma$ band below $E_F$, shifts down monotonically with the $x$ value and the local parameter bond angle, while the $\alpha_1$ band top crosses $E_F$ between $x = 0.2$ and 0.4. As illustrated in Fig. 7, the $T_c$ exhibits a non-linear behavior. It is natural to consider that the $\alpha_1$ band is responsible for the rapid increment in $T_c$, as it induces an abrupt change in the number of FSs from $x = 0.4$ to 0.2, as illustrated in Fig 7(b). In addition, the weight of the orbital character of $d_{xz}/d_{yz}$ of $\alpha_1$ band top decreases while $d_{xy}$ increases, as the $\alpha_1$ band sinks from above to below $E_F$. It has been theoretically predicted that the $\gamma$ hole FS pocket triggers a switch from high $T_c$ nodeless to low $T_c$ nodal gap symmetry in the IBSs [14–16], however, the present ARPES results indicate that the $\alpha_1$ band with $d_{xy}$ orbital character at the tip may play an important role. Recently, two different types of antiferromagnetic spin fluctuations (AFMSFs) were suggested for the LaFe(As$_{1-x}$P$_x$)(O$_{1-y}$F$_y$) [10]. The low-energy AFMSF at $x$ = 0.6 is caused by the nesting between FSs with $d_{xz}/d_{yz}$ orbital characters, while the high-energy AFMSF at $x$ = 0 is related to the FSs or bands with the orbital character of $d_{xz}/d_{yz}$ and $d_{xy}$. The bands around the Brillouin zone corner, which are out of the scope of current study, may influence the non-linear behavior of $T_c$ because the evolution of the FS shape is observed in the in-plane mapping illustrated in Fig. 1 (Schematic illustration in Fig. 7), i.e. from a propeller-like shape into a circular shape. As can be observed in the schematic illustration of the FS topology in Fig. 7(a), the nesting condition is good at $x$ = 0.6, as the orbital character and the size of FS between the inner hole pocket and electron pocket match with each other, provided the inner electron pocket is attributed to the $d_{xz}/d_{yz}$ orbital.

The FS nesting has been used to describe the origin of SC1 although the actual FS shape is inconsistent with the band calculation. In addition, newly discovered second SC dome in the hydrogen-doped La1111 further questions the FS nesting scenario. Among the P-free 1111 compounds, LaFeAsO$_{1-y}$H$_y$ is special because it has low and two-dome shaped $T_c$ as a function of $y$, while the other $Ln$FeAsO$_{1-y}$H$_y$ for $Ln$ = Ce, Sm, Gd solely exhibits a single dome feature with $T_c$ as high as 50 K [40]. SC3 is assigned to the special second SC dome (0.21 < $y$ < 0.53), with a maximum $T_c$ of 36 K at $y$ = 0.3. The two-domes in LaFeAsO$_{1-y}$H$_y$ merge into a single dome with a maximum $T_c$ = 52 K under pressure, thus suggesting that the merging of SC1 and SC3 is the main reason for the high $T_c$ [41]. Iimura et al. also predicted that a similar band splitting process (by the downward shift of anti-$d_{xy}$ band) occurs and forms an incipient band with tips constituted by $d_{xy}$ orbital in the SC3 regime [40]. In addition, their reported bond angle $\theta \sim 111°$ and calculated band structure for LaFeAsO$_{0.6}$H$_{0.4}$ are similar to those of our NdFeAsO$_{0.9}$F$_{0.1}$. In other words, the present ARPES results for $x$ = 0 may be related to the predicted SC3 band structure.

The diagonal electron hopping in the $d_{xy}$ orbital was recently suggested to play an important role in the SC3 regime in LaFeAsO$_{1-y}$H$_y$ due to the rapid decrease of $t_1$ compared to $t_2$ in the SC3 regime [42], where, $t_1$ and $t_2$ are the nearest and next nearest neighbor hopping between Fe sites in the real space. In the tight-binding picture, the energy difference between the $d_{xy}$ band at (0,0) and ($\pi,\pi$) in the unfolded Brillouin zone is approximately $8t_1$ [16]. Hence, we can estimate the $8t_1$ values by measuring the energy difference between the $d_{xy}$ $\gamma$ band and $d_{xy}$ orbital at the local maximum of $\alpha_1$ band. The energy difference is approximately 30 meV at $x$ = 0 and increases with increasing $x$ values. The increment is consistent with the theoretical prediction as the bond angle increases (pnictogen height decreases), where the indirect Fe-As-Fe hopping becomes dominant for the $t_1$ [42]. The present results indicate a very small value of $8t_1$, suggesting that NdFeAs$_{1-x}$P$_x$O$_{0.9}$F$_{0.1}$ with $x \sim$ 0 is located in the $t_2$ dominant region.

Interestingly, we also identified an enhancement in the intensity of the small flat $d_{xy}$ band when the temperature was cooled below $T_c$ for $x$ = 0 in our preliminary experiment. In fact, the flattening of $d_{xy}$ dispersion below $T_c$ in Fig. 2(g) is the evidence of the electron condensation. This further convinces us that the $d_{xy}$ at the incipient $\alpha_1$ band is involved in the superconductivity. A similar electron condensation on the incipient band has been reported recently [33,43]. The phonon boost effect is currently one of the major candidates for the source behind electron pairing in an incipient band [44,45]. Hesani et al. theoretically argued that a switching behavior from high to low $T_c$ originates from the orbital switching and that the $d_{xy}$ orbital character is a major ingredient to enhance $T_c$ [46]. The present findings strongly suggest that the energy shift of the $d_{xy}$ orbital character on the incipient $\alpha_1$ band plays a role in increasing $T_c$.

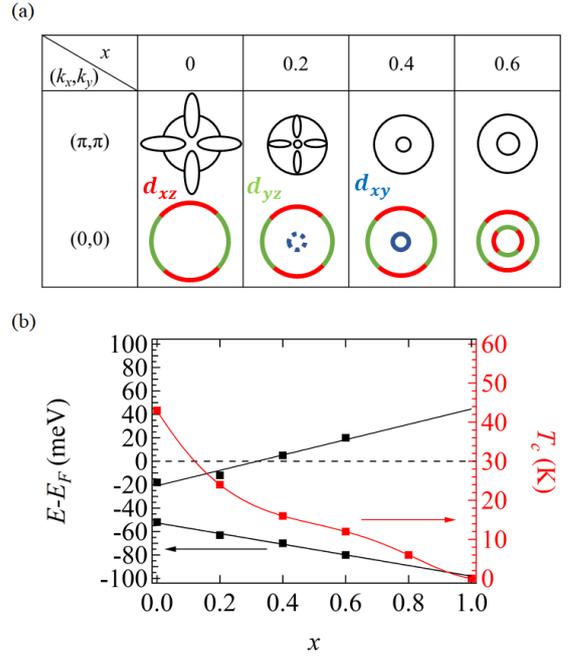

FIG. 7. (a) Schematic illustration of the Fermi surfaces in NdFeAs$_{1-x}$P$_x$O$_{0.9}$F$_{0.1}$ with $x$ = 0 – 0.6. The upper and lower panels illustrate the Fermi surfaces around the Brillouin zone corner and center, respectively. The orbital characters of Fermi surfaces are also presented for the lower panels. The switching of the orbital character from $d_{xz/yz}$ to $d_{XZ/YZ}$ is omitted for better visualization. (b) The $x$ dependence of $T_c$ (red dots) and the energy levels $\alpha_1$ and $\gamma$ band tops. The energy levels of $\alpha_1$ band top for $x$ = 0.4 and 0.6 are estimated by parabola-fitting the ARPES results.

In summary, we verified the existence of the $d_{xy}$ orbital character $\gamma$ band, which shifts toward the $E_F$ with decreasing P-content ($x$). With decreasing $x$ values, the degenerate $d_{xz}/d_{yz}$ $\alpha_1$ and $\alpha_2$ hole bands split. The $\alpha_1$ hole band sinks below $E_F$ at $x$ < 0.4, which triggers to an abrupt change in the Fermi surface topology, accompanied by a rapid enhancement of $T_c$. The presence of the $d_{xy}$ orbital character at the top of $\alpha_1$ hole band is likely to be an essential element for the pairing mechanism in the high-$T_c$ 1111-type IBSs as they are predicted for the SC3 with a regular tetrahedron structure. Our ARPES measurement on the NdFeAs$_{1-x}$P$_x$O$_{0.9}$F$_{0.1}$ reveals the systematic change in the band structure and elucidates the origin of the incipient band in the As-end 1111 compound.

Acknowledgment

The authors thank K. Kuroki and H. Usui for their helpful discussions. This research was supported by the Use-of-UVSOR Synchrotron Facility Program (Proposal No. 28-532, No. 28-823, No. 28-836, No. 29-533, No. 29-552, No. 29-843, No. 30-853, No. 19-561, No. 20-764, No.20-782) of the Institute for Molecular Science, Okazaki, Japan.

[1] Y. Kamihara, T. Watanabe, M. Hirano, and H. Hosono, J. Am. Chem. Soc. **130**, 3296 (2008).

[2] T. Nomura, S. W. Kim, Y. Kamihara, M. Hirano, P. V Sushko, K. Kato, M. Takata, A. L. Shluger, and H. Hosono, Supercond. Sci. Technol. **21**, 125028 (2008).

[3] H. Luetkens, H.-H. Klauss, M. Kraken, F. J. Litterst, T. Dellmann, R. Klingeler, C. Hess, R. Khasanov, A. Amato, C. Baines, M. Kosmala, O. J. Schumann, M. Braden, J. Hamann-Borrero, N. Leps, A. Kondrat, G. Behr, J. Werner, and B. Büchner, Nat. Mater. **8**, 305 (2009).

[4] S. Miyasaka, A. Takemori, T. Kobayashi, S. Suzuki, S. Saijo, and S. Tajima, J. Phys. Soc. Japan **82**, 124706 (2013).

[5] S. Matsuishi, T. Maruyama, S. Iimura, and H. Hosono, Phys. Rev. B **89**, 094510 (2014).

[6] K. T. Lai, A. Takemori, S. Miyasaka, F. Engetsu, H. Mukuda, and S. Tajima, Phys. Rev. B **90**, 064504 (2014).

[7] S. Miyasaka, M. Uekubo, H. Tsuji, M. Nakajima, S. Tajima, T. Shiota, H. Mukuda, H. Sagayama, H. Nakao, R. Kumai, and Y. Murakami, Phys. Rev. B **95**, 214515 (2017).

[8] T. Kawashima, S. Miyasaka, H. Tsuji, T. Yamamoto, M. Uekubo, A. Takemori, K. T. Lai, and S. Tajima, Sci. Rep. **11**, 10006 (2021).

[9] C. Shen, B. Si, C. Cao, X. Yang, J. Bao, Q. Tao, Y. Li, G. Cao, and Z. A. Xu, J. Appl. Phys. **119**, 083903 (2016).

[10] T. Shiota, H. Mukuda, M. Uekubo, F. Engetsu, M. Yashima, Y. Kitaoka, K. T. Lai, H. Usui, K. Kuroki, S. Miyasaka, and S. Tajima, J. Phys. Soc. Japan **85**, 053706 (2016).

[11] A. Takemori, T. Hajiri, S. Miyasaka, Z. H. Tin, T. Adachi, S. Ideta, K. Tanaka, M. Matsunami, and S. Tajima, Phys. Rev. B **98**, 100501 (2018).

[12] V. Vildosola, L. Pourovskii, R. Arita, S. Biermann, and A. Georges, Phys. Rev. B **78**, 064518 (2008).

[13] H. Usui, K. Suzuki, and K. Kuroki, Sci. Rep. **5**, 11399 (2015).

[14] I. I. Mazin, D. J. Singh, M. D. Johannes, and M. H. Du, Phys. Rev. Lett. **101**, 057003 (2008).

[15] K. Kuroki, S. Onari, R. Arita, H. Usui, Y. Tanaka, H. Kontani, and H. Aoki, Phys. Rev. Lett. **101**, 087004 (2008).

[16] K. Kuroki, H. Usui, S. Onari, R. Arita, and H. Aoki, Phys. Rev. B **79**, 224511 (2009).

[17] T. Misawa, K. Nakamura, and M. Imada, Phys. Rev. Lett. **108**, 177007 (2012).

[18] L. de' Medici, G. Giovannetti, and M. Capone, Phys. Rev. Lett. **112**, 177001 (2014).

[19] T. Misawa and M. Imada, Nat. Commun. **5**, 5738 (2014).

[20] M. Yi, Y. Zhang, Z. X. Shen, and D. Lu, npj Quantum Mater. **2**, (2017).

[21] M. Yi, H. Pfau, Y. Zhang, Y. He, H. Wu, T. Chen, Z. R. Ye, M. Hashimoto, R. Yu, Q. Si, D. H. Lee, P. Dai, Z. X. Shen, D. H. Lu, and R. J. Birgeneau, Phys. Rev. X **91**, 041049 (2019).

[22] M. D. Watson, P. Dudin, L. C. Rhodes, D. V. Evtushinsky, H. Iwasawa, S. Aswartham, S. Wurmehl, B. Büchner, M. Hoesch, and T. K. Kim, npj Quantum Mater. **4**, 36 (2019).

[23] H. Pfau, C. R. Rotundu, J. C. Palmstrom, S. D. Chen, M. Hashimoto, D. Lu, A. F. Kemper, I. R. Fisher, and Z.-X. Shen, Phys. Rev. B **99**, 035118 (2019).

[24] S.-I. Kimura, T. Ito, M. Sakai, E. Nakamura, N. Kondo, T. Horigome, K. Hayashi, M. Hosaka, M. Katoh, T. Goto, T. Ejima, and K. Soda, Rev. Sci. Instrum. **81**, 53104 (2010).

[25] C. Liu, Y. Lee, A. D. Palczewski, J.-Q. Yan, T. Kondo, B. N. Harmon, R. W. McCallum, T. A. Lograsso, and A. Kaminski, Phys. Rev. B **82**, 075135 (2010).

[26] H. Eschrig, A. Lankau, and K. Koepernik, Phys. Rev. B **81**, 155447 (2010).

[27] L. X. Yang, B. P. Xie, Y. Zhang, C. He, Q. Q. Ge, X. F. Wang, X. H. Chen, M. Arita, J. Jiang, K. Shimada, M. Taniguchi, I. Vobornik, G. Rossi, J. P. Hu, D. H. Lu, Z. X. Shen, Z. Y. Lu, and D. L. Feng, Phys. Rev. B **82**, 104519 (2010).

[28] A. Damascelli, Z. Hussain, and Z.-X. Shen, Rev. Mod. Phys. **75**, 473 (2003).

[29] Y. Zhang, F. Chen, C. He, B. Zhou, B. P. Xie, C. Fang, W. F. Tsai, X. H. Chen, H. Hayashi, J. Jiang, H. Iwasawa, K. Shimada, H. Namatame, M. Taniguchi, J. P. Hu, and D. L. Feng, Phys. Rev. B **83**, 054510 (2011).

[30] B. Mansart, V. Brouet, E. Papalazarou, M. Fuglsang Jensen, L. Petaccia, S. Gorovikov, A. N. Grum-Grzhimailo, F. Rullier-Albenque, A. Forget, D. Colson, and M. Marsi, Phys. Rev. B **83**, 064516 (2011).

[31] Y. Zhang, C. He, Z. R. Ye, J. Jiang, F. Chen, M. Xu, Q. Q. Ge, B. P. Xie, J. Wei, M. Aeschlimann, X. Y. Cui, M. Shi, J. P. Hu, and D. L. Feng, Phys. Rev. B **85**, 085121 (2012).

[32] I. Nishi, M. Ishikado, S. Ideta, W. Malaeb, T. Yoshida, A. Fujimori, Y. Kotani, M. Kubota, K. Ono, M. Yi, D. H. Lu, R. Moore, Z.-X. Shen, A. Iyo, K. Kihou, H. Kito, H. Eisaki, S. Shamoto, and R. Arita, Phys. Rev. B **84**, 014504 (2011).

[33] A. Charnukha, D. V Evtushinsky, C. E. Matt, N. Xu, M. Shi, B. Büchner, N. D. Zhigadlo, B. Batlogg, and S. V Borisenko, Sci. Rep. **5**, 18273 (2015).

[34] H. Usui, K. Suzuki, and K. Kuroki, Supercond. Sci. Technol. **25**, 084004 (2012).

[35] See Supplemental Material for the further analysis of ARPES results.

[36] A. Charnukha, S. Thirupathaiah, V. B. Zabolotnyy, B. Büchner, N. D. Zhigadlo, B. Batlogg, A. N. Yaresko, and S. V Borisenko, Sci. Rep. **5**, 10392 (2015).

[37] P. Zhang, J. Ma, T. Qian, Y. G. Shi, A. V Fedorov, J. D. Denlinger, X. X. Wu, J. P. Hu, P. Richard, and H. Ding, Phys. Rev. B **94**, 104517 (2016).

[38] H. Usui and K. Kuroki, Phys. Rev. B **84**, 024505 (2011).

[39] H. Usui, K. Suzuki, K. Kuroki, N. Takeshita, P. M. Shirage, H. Eisaki, and A. Iyo, Phys. Rev. B **87**, 174528 (2013).

[40] S. Iimura, S. Matuishi, H. Sato, T. Hanna, Y. Muraba, S. W. Kim, J. E. Kim, M. Takata, and H. Hosono, Nat. Commun. **3**, 943 (2012).

[41] H. Takahashi, H. Soeda, M. Nukii, C. Kawashima, T. Nakanishi, S. Iimura, Y. Muraba, S. Matsuishi, and H. Hosono, Sci. Rep. **5**, 7829 (2015).

[42] K. Suzuki, H. Usui, S. Iimura, Y. Sato, S. Matsuishi, H. Hosono, and K. Kuroki, Phys. Rev. Lett. **113**, 027002 (2014).

[43] H. Miao, T. Qian, X. Shi, P. Richard, T. K. Kim, M. Hoesch, L. Y. Xing, X.-C. Wang, C.-Q. Jin, J.-P. Hu, and H. Ding, Nat. Commun. **6**, 6056 (2015).

[44] X. Chen, S. Maiti, A. Linscheid, and P. J. Hirschfeld, Phys. Rev. B **92**, 224514 (2015).

[45] Y. Bang, Sci. Rep. **9**, 3907 (2019).

[46] M. Hesani and A. Yazdani, Physica C **553**, 38 (2018).